\documentclass[fleqn,usenatbib]{mnras}
\usepackage{amssymb}
\usepackage{mathrsfs}
\usepackage{xspace}

\usepackage[T1]{fontenc}
\usepackage{ae,aecompl}

\newcommand{\rsun}{\,\mbox{$\rm R_{\odot}$}\xspace}

\newcommand{\msun}{\,\mbox{$\rm M_{\odot}$}\xspace}
\newcommand{\mearth}{\,\mbox{$\rm M_{\oplus}$}\xspace}

\newcommand{\ionjb}[2]{#1{~\sc{\romannumeral #2}}}

\newcommand{\kms}{\hbox{km~s$^{-1}$}\xspace}
\newcommand{\ms}{\hbox{ms$^{-1}$}\xspace}

\newcommand{\vsini}{\hbox{$v$\,sin\,$i$}\xspace}

\newcommand{\degs}{$\degr$\xspace}
\newcommand{\chisq}{$\chi^{2}$\xspace}

\newcommand{\ha}{H$\alpha$\xspace}

\newcommand{\dotstom}{{\sc DoTS}\xspace}

\newcommand{\harps}{{\sc harps}\xspace}

\newcommand{\lp}{\hbox{LP 944-20}\xspace}

\newcommand{\irec}{$i_{\rm rec}$\xspace}

\newcommand{\rstar}{$R_*$\xspace}

\newcommand{\kstar}{$K_*$\xspace}

\newcommand{\prot}{$P_{\rm rot}$\xspace}
\newcommand{\porb}{$P_{\rm orb}$\xspace}

\newcommand{\mstar}{$M_*$\xspace}
\newcommand{\mplan}{$M_{\rm plan}$\xspace}

\newcommand{\rvsim}{RV$_{\rm sim}$\xspace}
\newcommand{\rvrec}{RV$_{\rm rec}$\xspace}
\newcommand{\rvcorr}{RV$_{\rm corr}$\xspace}

\newcommand{\snrobs}{S/N$_{\rm obs}$\xspace}
\newcommand{\snrdecon}{S/N$_{\rm decon}$\xspace}

\newcommand{\nobs}{$N_{\rm obs}$\xspace}

\usepackage{graphicx}	
\usepackage{amsmath}	
\usepackage{amssymb}	

\title[Removing starspot activity]{Recovering planet radial velocity signals in the presence of starspot activity in fully convective stars}

\author[J.R.~Barnes et al.]
{J.R.~Barnes$^{1}$, 
S.V. Jeffers$^{2}$,
G.~Anglada-Escud\'{e}$^{3}$,
C.A.~Haswell$^{1}$, 
H.R.A.~Jones$^{4}$, \newauthor
M. Tuomi$^{4}$,
F. Feng$^{4}$,
J.S.~Jenkins$^{5}$, 
P.~Petit$^{6,7}$ \\
$^{1}$ Department of Physical Sciences, The Open University, Walton Hall, Milton Keynes MK7 6AA, UK. \\
$^{2}$ Institut f\"{u}r Astrophysik, Georg-August-Universit\"{a}t, Friedrich-Hund-Platz 1, Friedrich-Hund-Platz 1, D-37077 G\"{o}ttingen. Germany. \\
$^{3}$ School of Physics and Astronomy, Queen Mary, University of London, 327 Mile End Rd. London, UK \\
$^{4}$ Centre for Astrophysics Research, University of Hertfordshire, College Lane, Hatfield AL10 9AB, UK \\
$^{5}$ Departamento de Astronom\'{i}a, Universidad de Chile, Camino del Observatorio 1515, Las Condes, Santiago. Chile. \\
$^{6}$ Universit\'{e} de Toulouse, UPS-OMP, Institut de Recherche en Astrophysique et Plan\'{e}tologie, 31000 Toulouse, France. \\
$^{7}$ CNRS, Institut de Recherche en Astrophysique et Plan\'{e}tologie, 14 Avenue \'{E}douard Belin, 31400 Toulouse, France. \\
}

\date{Accepted for publication in MNRAS (2016 December 2)}

\pubyear{201X}

\begin{document}

\label{firstpage}
\pagerange{\pageref{firstpage}--\pageref{lastpage}}
\maketitle

\begin{abstract}
Accounting for stellar activity is a crucial component of the search for ever-smaller planets orbiting stars of all spectral types. We use Doppler imaging methods to demonstrate that starspot induced radial velocity variability can be effectively reduced for moderately rotating, fully convective stars. Using starspot distributions extrapolated from sunspot observations, we adopt typical M dwarf starspot distributions with low contrast spots to synthesise line profile distortions. The distortions are recovered using maximum entropy regularised fitting and the corresponding stellar radial velocities are measured. The procedure is demonstrated for a late-M star harbouring an orbiting planet in the habitable zone. The technique is effective for stars with \hbox{\vsini $=$} \hbox{$1$\,-\,$10$ \kms}, reducing the stellar noise contribution by factors of nearly an order of magnitude. With a carefully chosen observing strategy, the technique can be used to determine the stellar rotation period and is robust to uncertainties such as unknown stellar inclination. While demonstrated for late-type M stars, the procedure is applicable to all spectral types.
\end{abstract}




\section{Introduction}
\protect\label{section:intro}
Efforts aimed at determining planet occurrence rates are now focusing on low mass planets orbiting the lowest mass stars.
Instrumental precision, which is limited by stability, and intrinsic stellar activity are the main factors that determine the number of observations needed to reliably recover planetary signals. The most stable spectrometers amongst the current generation of dedicated radial velocity (RV) instruments regularly achieve \hbox{$\sim1$ \ms}\ \citep{mayor03harps,cosentino12harpsn}.
However, obtaining this level of precision outside the standard optical range ($0.4\,-\,0.7$ \micron), at red-optical ($0.6\,-\,1.0$ \micron) and near infrared ($> 1$ \micron) wavelengths is also required if the lowest mass, fully convective stars, with \mstar $< 0.35$ \msun (spectral type M4V or later), are to be surveyed efficiently. \citet{barnes14rops} demonstrated precision down to \hbox{$\sim$\,$2.5$ \ms} is possible with current technology operating at red-optical wavelengths, and \citet{gao16cshell} have demonstrated similar precision down to $\sim$\,$2$ \ms in the near infrared K band. There are also dedicated RV surveys, including the Habitable Zone Planet Finder \citep{mahadevan14hzp} and SpectroPolarim\`{e}tre Infra-Rouge (SPIRou) \citep{thibault12spirou} that will target the lowest mass stars at the bottom of the main sequence with 1 \ms instrumental RV precision. For M dwarfs, although the red-optical contains more Doppler information in the thousands of available absorption lines, enabling greater radial velocity precision to be achieved \citep{reiners10rvs}, the spectral energy distribution of low mass stars peaks at near infrared wavelengths. The Calar Alto high-Resolution search for M dwarfs with Exoearths with Near-infrared and optical \'{E}chelle Spectrographs survey ({CARMENES}) \citep{quirrenbach10carmenes} is addressing this issue by covering both spectral regions simultaneously. 

Despite the expectation that fully convective stars cannot possess a solar-like dynamo in the absence of a convective-radiation boundary where strong shearing is believed to take place, M stars are still observed to possess strong magnetic fields. Moreover, by mid-M, field stars in the solar neighbourhood have not spun down and typically possess significant rotation \citep{jenkins09mdwarfs}; for an M5V star, the mean equatorial rotation, \vsini $\sim 6$ \kms. { \citet{newton16mearth} have however suggested that \vsini upper limits measured from low resolution (R $< 40,000$) spectra significantly overestimate rotation rates. By measuring rotation periods in a sample of nearly 400 M stars, they find that the mid-low mass M dwarfs tend to exhibit short rotation periods of $P < 10$\ d or periods of $P > 70$ d, with a dearth of intermediate rotators. They suggest that this dichotomy exists because stars maintain fast rotation for a few Gyr before rapid spin down to slow rotation (~100 d) by an age of 5 Gyr.}

Fully convective stars show activity all the way down to spectral type M9V, despite an overall decline in activity \citep{mohanty03activity,reiners10activity}. There is a clear correlation between rotation period and \ionjb{Ca}{2} chromospheric emission index, \hbox{log\ $R'_{HK}$}, down to mid-M \citep{suarez-mascerno2015}, while for 15 M5-M9 dwarfs, \citet{barnes14rops} found a correlation between r.m.s. RV and \ha emission strength. Doppler imaging studies by \citet{morin08v374peg,phanbao09} and \citet{barnes15mdwarfs} have shown that line profile variability in fully convective stars can be interpreted as starspots. This interpretation is valid, even amongst the latest spectral types \citep{barnes15mdwarfs}, where \hbox{$\sim100$ \ms}\ RV { variation} is observed for the rapidly rotating M9 dwarf, \lp (\vsini = 31 \kms). Fully convective field M dwarfs are thus moderate rotators that display activity which causes significant RV { variations}. Crucially, monitoring only slow rotators amongst a population of moderate rotators will favour low-inclination (pole-on) stars for which planet detection is less favoured. If we wish to conduct unbiased RV surveys to determine planet occurrence rates for fully convective stellar hosts, we must find a means to account for the stellar activity component.

Modelling and removal of the stellar activity component in RV measurements, which can swamp planetary signals, can be achieved in a number of ways.
Traditional techniques involving measuring the line bisector are most effective for simple cases where a single spot, or spot group, on the star induces the line asymmetries. Other methods have recently been developed that enable a more sophisticated approach that identifies the stellar activity signatures directly from the line profiles rather than the RV measurement. By fitting Gaussian profiles to residual time series spectra \cite{moulds13} was able to reduce starspot induced RVs by more than 80 per cent. The technique is applicable to stars with significant rotation: i.e. \vsini\ = 10\,-\,50 \kms. \cite{dumusque14soap2} has shown that RV jitter can be removed by modelling activity regions on the stellar surface, but this has only been applied to stars with spectral type similar to the Sun. \cite{petit15planets} have applied maximum-entropy techniques to simultaneously recover planetary RVs with spot signatures on stars. The method was shown to work well for \vsini\ values of $\geq 20$ \kms. Below this \vsini it is unclear whether Doppler imaging techniques can simultaneously recover the planet and spot distribution.

A large number of variables must be considered for the detection of a planetary RV signature in the presence of stellar activity. Planetary orbital elements include the orbital period, \porb, and the stellar reflex motion velocity amplitude of the star, \kstar. The stellar activity includes a potentially unknown distribution of spots with some characteristic photospheric-spot contrast that distort the line profiles in a { quasi-}periodic manner, associated with the stellar rotation period, \prot. The observation strategy is important for recovery of the planet and modelling of the spots. Specifically good phase sampling on timescales that minimises activity evolution and our ability to obtain spectra with sufficient S/N ratios will determine the effectiveness of our ability to model the line profiles. In this paper, we apply Doppler imaging techniques to show that the effects of cool starspots on line profiles can be modelled for stars with \hbox{\vsini $=$} \hbox{1\,-\,10 \kms}. We then show this enables effective reduction of starspot induced radial velocities. Reconstruction of the line profiles using maximum entropy regularisation has the advantage that no prior assumption of the spot locations and sizes is required. In \S \ref{section:technique}, we describe the spot models and line profile simulation and recovery. The technique is illustrated in \S \ref{section:sims} using a representative case. We also investigate the ability of the technique to recover parameters, including stellar rotation period and the importance of stellar axial inclination. Sensitivities for a broader set of cases are explored in \S \ref{section:sensitivities} before a brief summary and discussion in \S \ref{section:discussion}.

\begin{figure}
\begin{center}
\begin{tabular}{ll}
\hspace{0mm}
\includegraphics[width=39mm,angle=0]{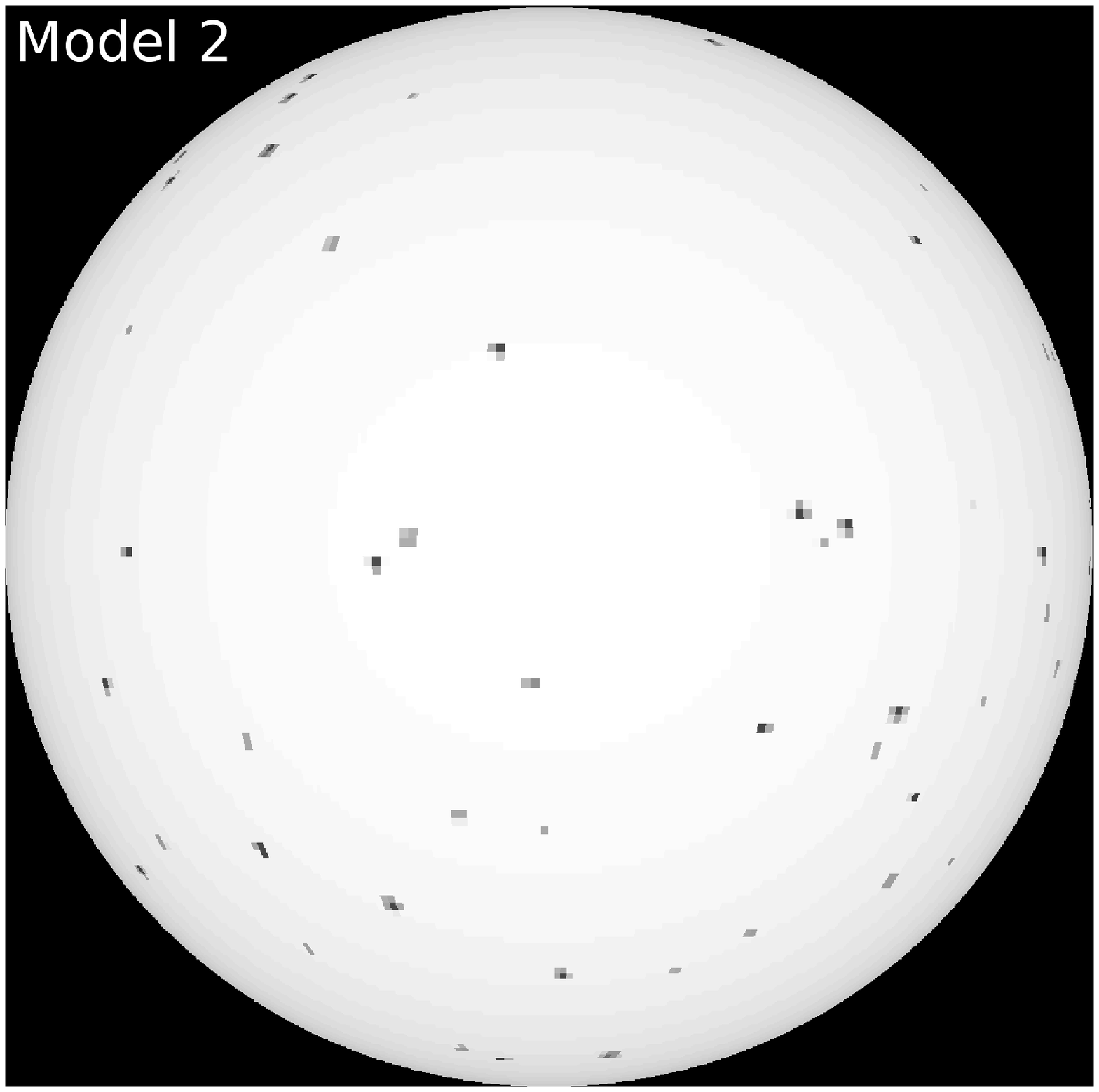}  & 
\hspace{-3mm}
\includegraphics[width=39mm,angle=0]{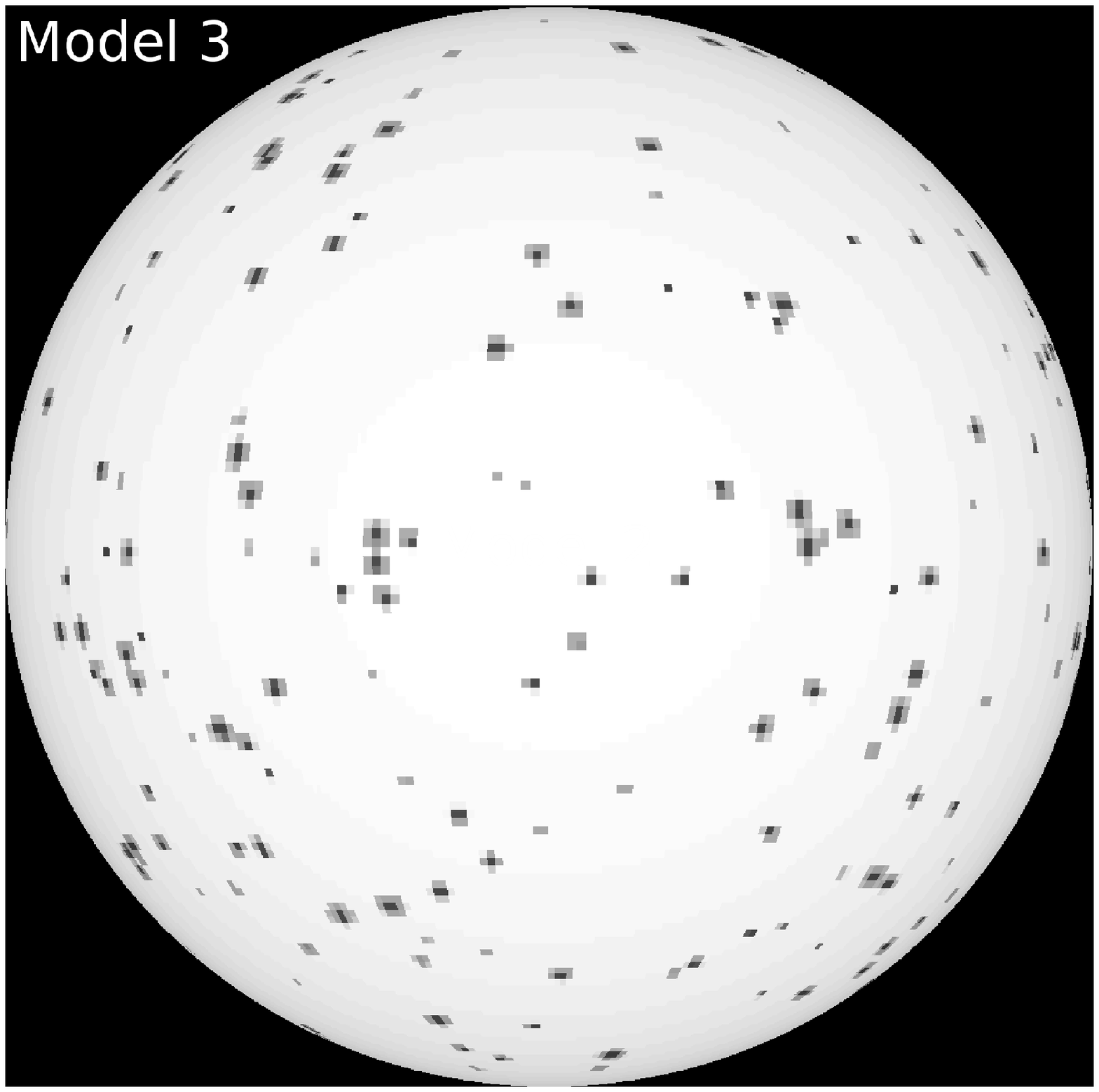}  \\
\end{tabular}
\begin{tabular}{c}
\hspace{0mm}
\includegraphics[width=39mm,angle=0]{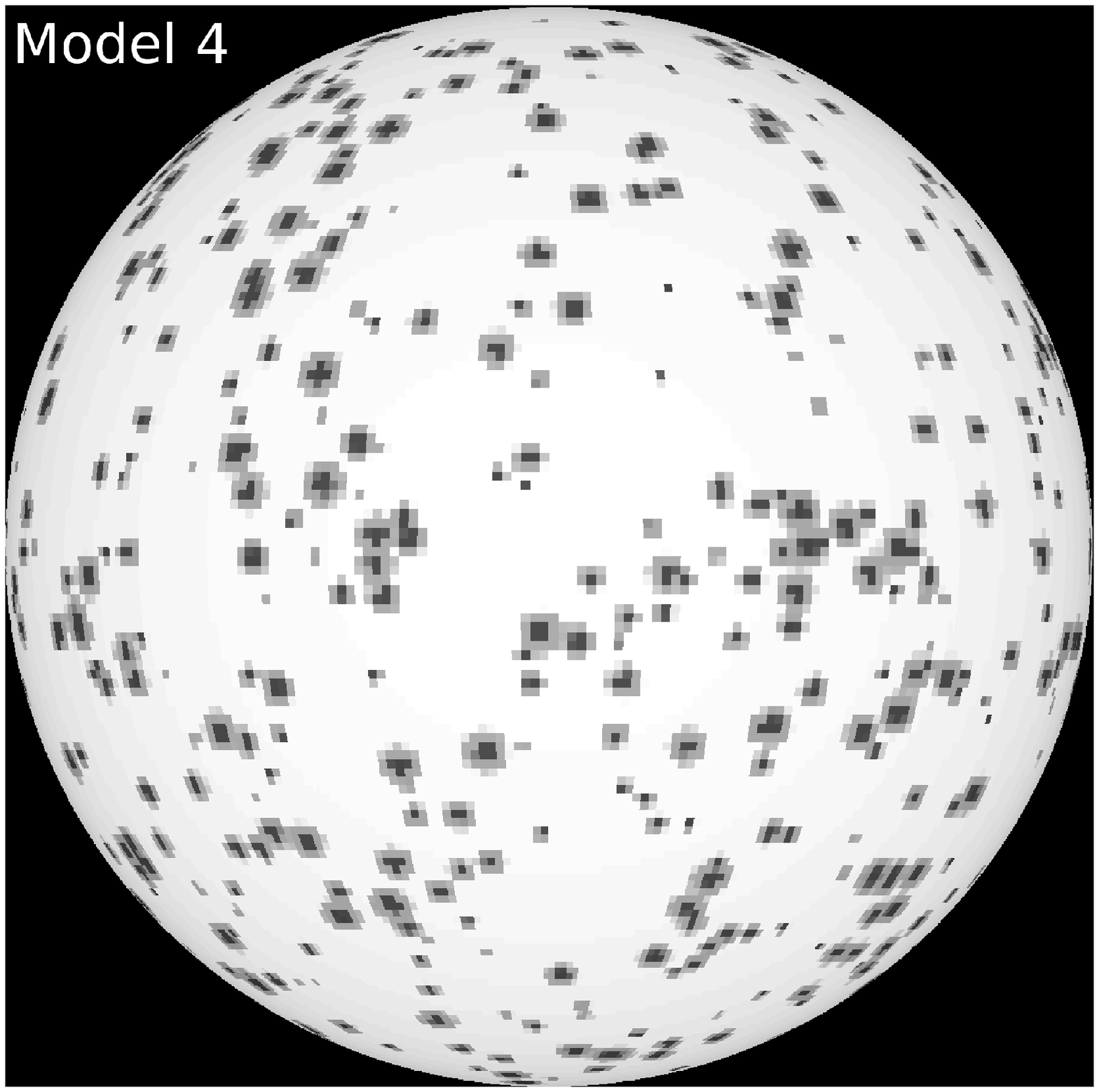}  \\ 
\end{tabular}
\end{center}   
\vspace{5mm} 
\caption{Spot Models 2 to 4 based on \citet{jeffers05activelongs} and adapted to M dwarf simulations in \citet{barnes11jitter}. Model 2 is analogous to solar max activity level while Model 3 represents a high solar activity case. The respective mean spot filling factors are 0.3, 1.9 and 9.0 per cent.}
\protect\label{fig:spotims}
\end{figure}

\section{Technique}
\protect\label{section:technique}

\subsection{Spot Models}
\protect\label{section:spotmodels}
In \cite{barnes11jitter}, we used extrapolated solar models to generate spot size distributions for greater than solar activity levels, following the work of \citet{jeffers05activelongs}. A key difference of our models compared with solar spot observations is that rather than confining spots to low-latitudes, we allowed spots to be located at all latitudes and longitudes. Evidence for spots distributed more uniformly than FGK dwarf stars comes from previous findings from Doppler imaging studies of early M stars \citep{barnes01mdwarfs,barnes04hkaqr} and more recent studies of stars at or below the fully convective boundary \citep{morin08v374peg,phanbao09,barnes15mdwarfs,barnes16zenodo}. The recovered spot distributions, along with the low contrasts required to model the spots, confirmed the low photometric amplitude variability observed during earlier monitoring of the latest M dwarfs \citep{rockenfeller06mdwarfs} { and in the much larger MEarth Project sample \citep{newton16mearth}}. 

Figure \ref{fig:spotims} shows the three starspot models that we use in this paper. These models are identical to those presented in \citet{barnes11jitter} and are based on the models of \citet{jeffers05activelongs}, but also include umbral and penumbral regions. We retain the same model numbering for consistency: Model 2 represents solar maximum activity, while Models 3 and 4 are for extrapolated activity levels with higher degrees of spot coverage. 

\subsection{Line profile modelling and RV recovery}
\protect\label{section:profmodel}

We used our spot models to investigate the effect of activity on precision radial velocity measurements of M dwarfs. We calculated synthetic line profiles using spotted 3D stellar models and calculated the resulting RVs by cross-correlating the profiles. We tested our ability to recover planet-induced RVs by combining our simulated spot-induced RVs with various planetary RV signals. In addition to investigating the effect of spot induced jitter on M dwarfs, we have also investigated the effect of stellar activity on rapidly rotating G and K dwarfs \citep{jeffers14activity}. { These simulations incorporated the effects of convective blueshift and modelled facular regions} in addition to starspot activity. 

Here we restrict simulations to starspots alone in order to demonstrate the feasibility of removing spot induced distortions in slowly rotating stars. Doppler Tomography of Stars (\dotstom) is a Doppler imaging code that utilises a two-temperature model to recover surface brightness distributions of active stars \citep{cameron01mapping}. \dotstom uses maximum entropy regularisation to obtain image solutions that minimise artefacts in the presence of noise; it has been used extensively to recover spot distributions on single stars and binaries. A brief overview has also been given recently in \cite{barnes15mdwarfs}, with application in particular to late M dwarfs. For an image with $i$ pixels, spot filling factors $f_i$ are obtained with \dotstom, assuming a two-temperature model representing the spots and photosphere. For stars that are rotating slowly and for which insufficient Doppler resolution elements can be obtained across the stellar line profile, the derived images are not particularly informative. The regularised fitted line profiles nevertheless offer the potential for distinguishing between distortion due to spots and line centroid shifts due to an orbiting extrasolar planet.

We cross-correlate the fittted line profiles recovered with \dotstom, but crucially, do not incorporate the small RVs induced by the planet in the recovery procedure. Similarly, we cross-correlate the input synthetic profiles, { which contain both the spot induced RV variations} and the planet induced RV { variations}. Any correlation between the simulated and recovered RV can be attributed to stellar activity modulated at the stellar rotation period and can be subtracted from the observed RVs. This approach will be most effective when the \prot and \porb are distinct, but further analysis of line profile moments \citep{berdinas16harps,anglada16proxima} may be useful for distinguishing more difficult cases. Similarly, the method is applicable when starspots are the main contributor to line profile {\em shape} variability above the noise level, since the model is designed to fit absorption line distortions (due to spots) and not line shifts (due to an orbiting planet).

\begin{table}
\begin{center}
\caption{System parameters for simulation based on the Proxima \newline \hbox{Centauri b} system, but with a higher \vsini and thus \prot. \label{tab:params}}
\begin{tabular}{lcc}
\hline
                              & Star   &  \\
\hline                       
\vsini\                       & [\kms]   & 5 \\
Stellar radius [\rstar]       & [\rsun]  & 0.14 \\
Rotation period [\prot]       & [d]      & 1.23 \\
Axial inclination [$i$]       & [degs]   & 60 \\
$T_{\rm phot}$                & [K]      & 3000 \\
$T_{\rm spot}$                & [K]      & 2700 \\
\hline                       
                              & Planet   &  \\
\hline                       
Orbital period [\porb]        & [d]     & 11.2 \\
Stellar reflex ampl. [\kstar] & [\ms]   & 2 \\                  
\hline
\end{tabular}
\end{center}
\end{table}


\begin{figure*}
\begin{center}
\includegraphics[width=68mm,height=160mm,angle=270]{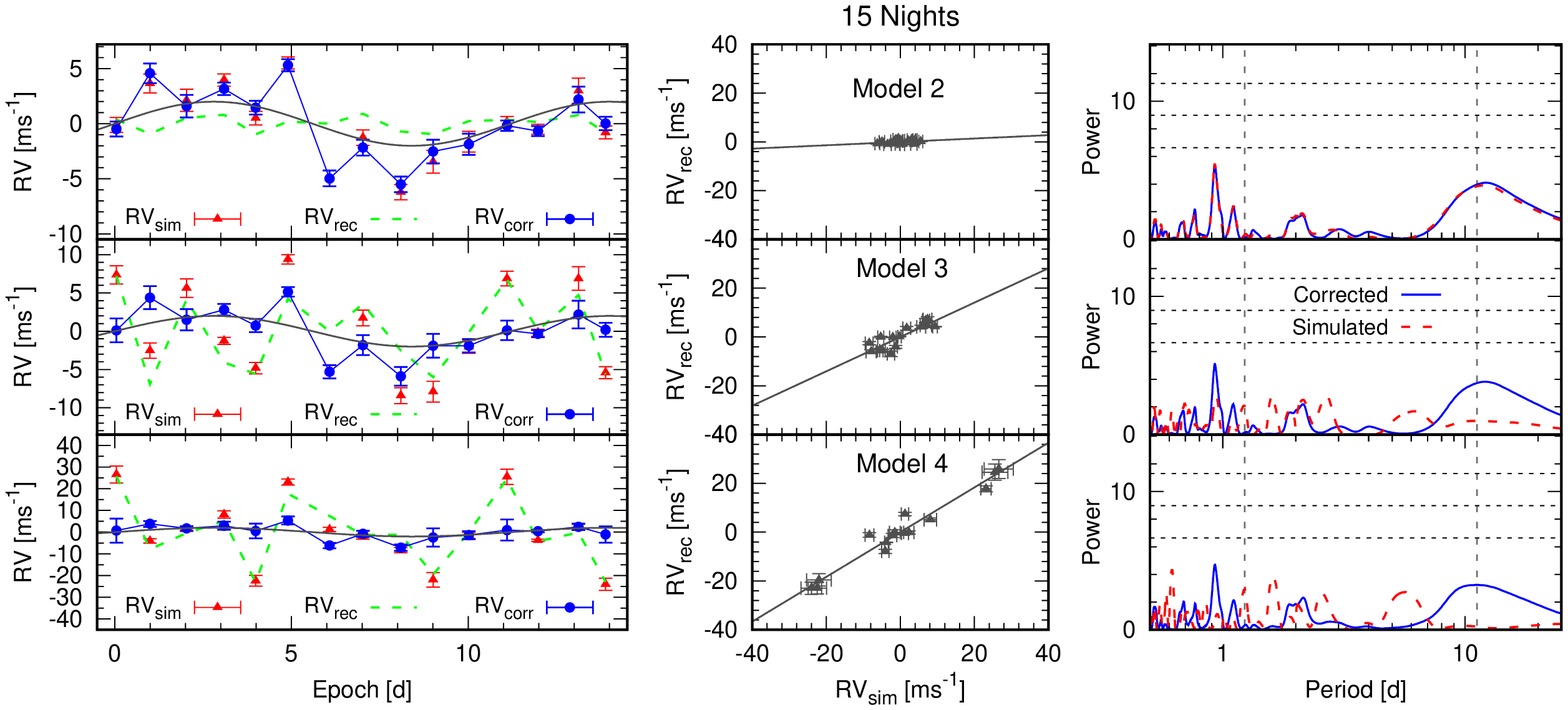} \\
\vspace{2mm}
\includegraphics[width=68mm,height=160mm,angle=270]{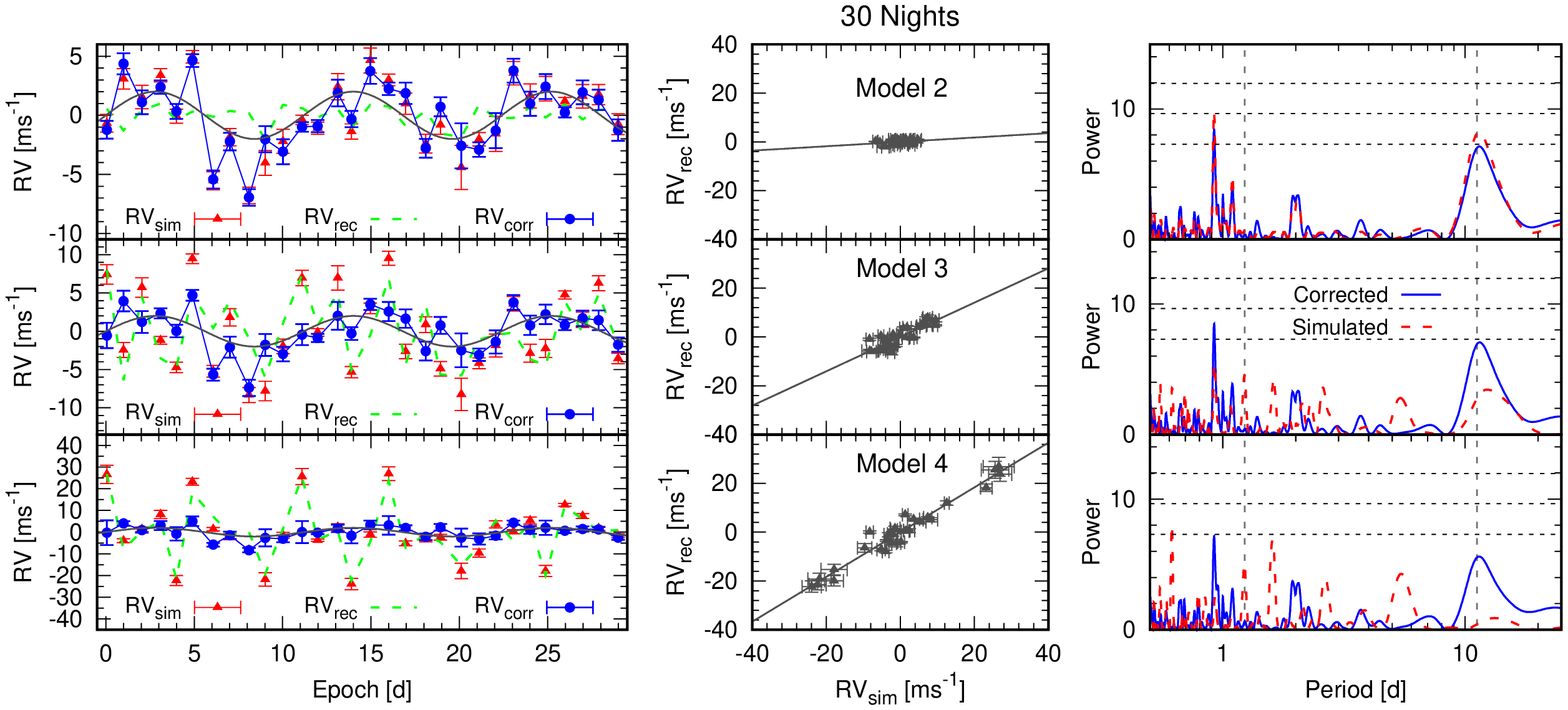} \\
\vspace{2mm}
\includegraphics[width=68mm,height=160mm,angle=270]{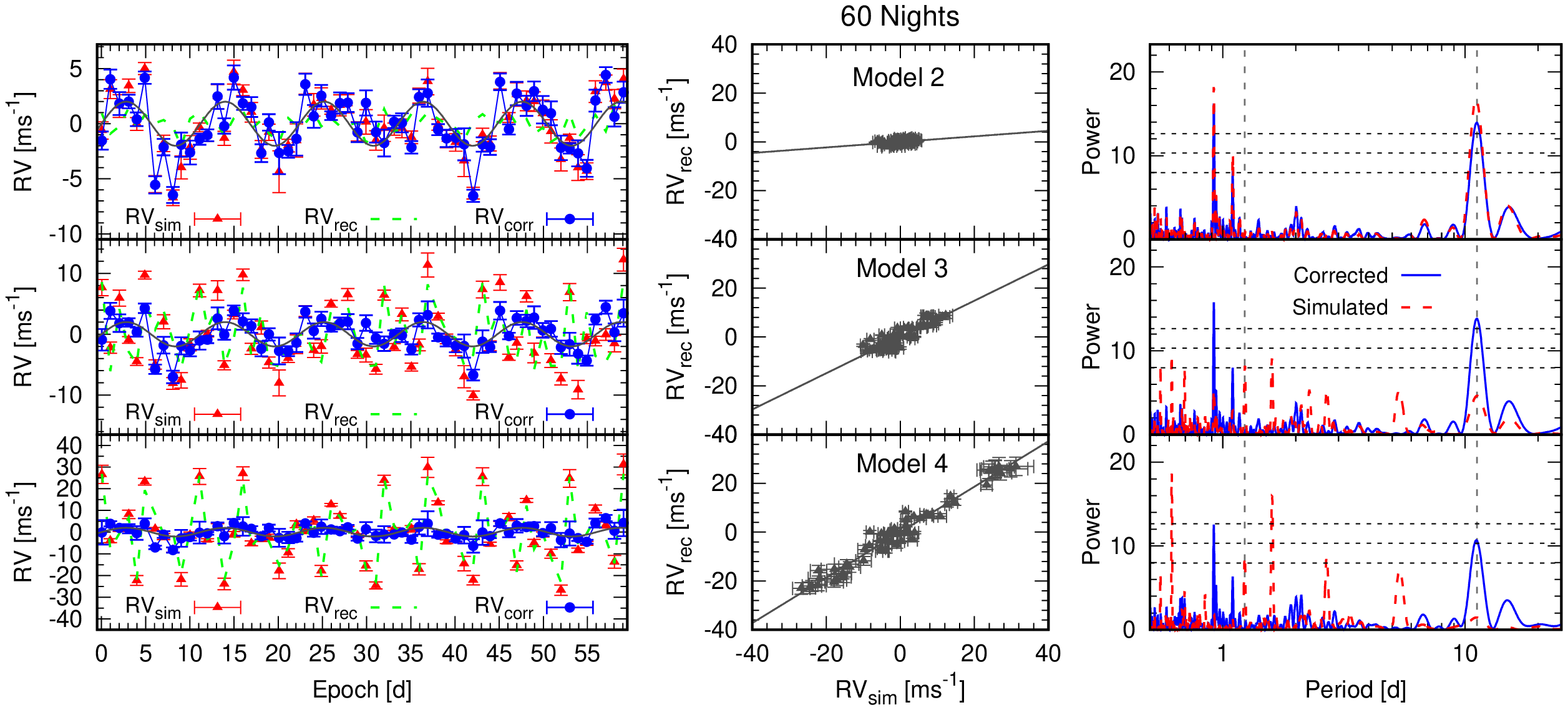} \\
\end{center}
\caption{Simulated RV curves comprising a \porb $= 11.2$ d planet that induces a stellar semi-amplitude of \kstar $ = 2$ \ms in the presence of stellar modulations from starspot Models 2\,-\,4 (see Figure 1). Stellar parameters are \vsini = 5 \kms, \prot $= 1.23$ d and $i = 60$\degs. Simulations for \nobs = 15 (top), 30 (middle) and 60 nights (bottom) are plotted. Left: The simulated spot + planet RVs and maximum entropy recovered RVs are respectively denoted \rvsim and \rvrec. The corrected radial velocities are \rvcorr = \rvsim - \rvrec. The model planetary RVs are shown by the (grey) sine curve. Centre: Plots showing the correlation between \rvrec and \rvsim. Right: Lomb-Scargle periodograms for \rvsim (Simulated) and { \rvcorr} (Corrected) RVs. Dashed vertical lines indicate \prot and \porb. False alarm probability levels, FAP = 0.001, 0.01 and 1 (i.e. 0.1, 1 and 10 per cent) are indicated by the horizontal lines (top to bottom respectively).}
\protect\label{fig:simulations1}
\end{figure*}

\begin{figure}
\begin{center}
\includegraphics[height=84.5mm,angle=270]{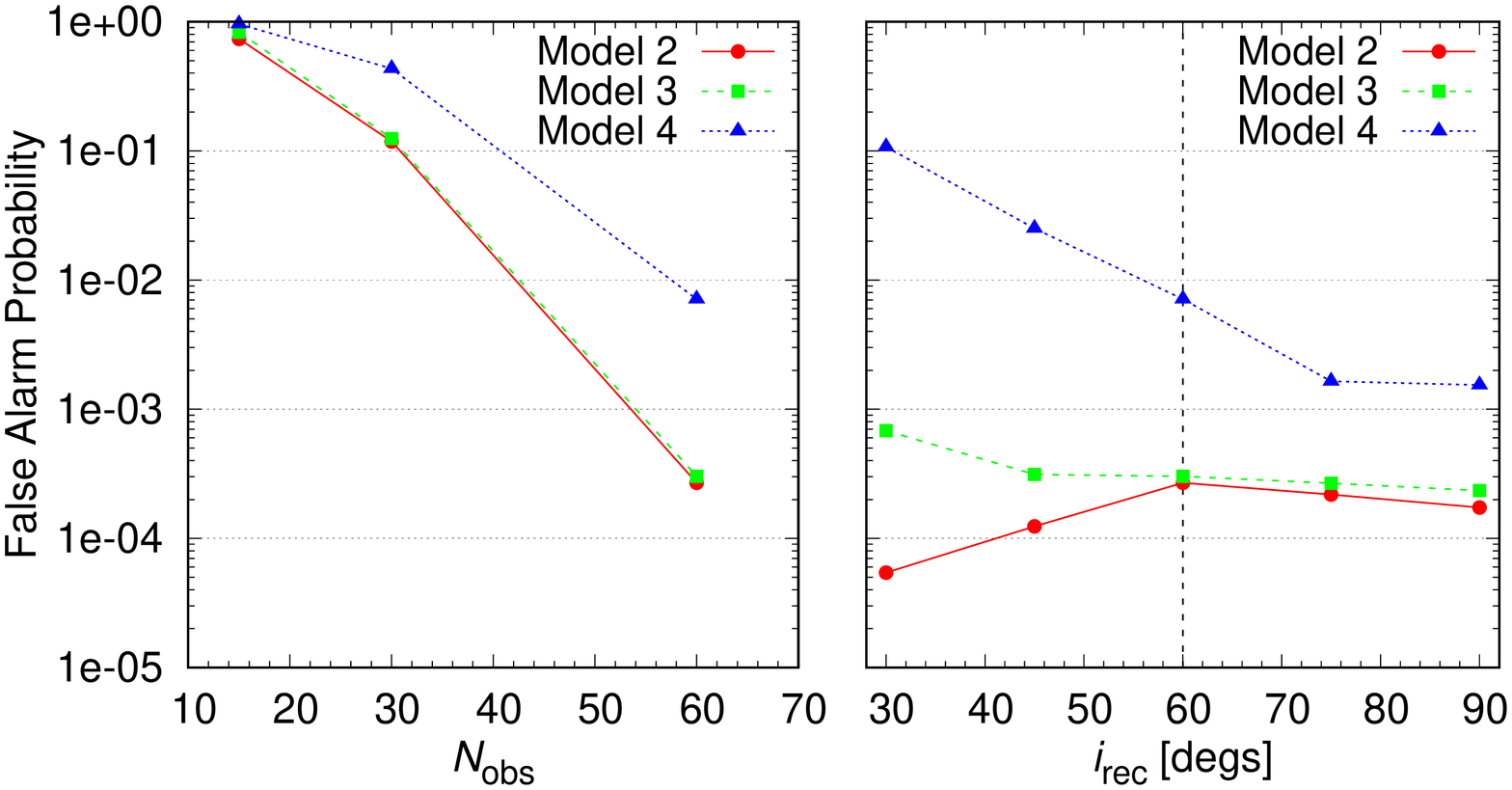}  \\
\caption{False alarm probabilities (FAP) of \porb for Models 2\,-\,4 after removing the modelled starspot contribution. Left: FAP vs number of observations, \nobs. Right: For the \nobs = 60 case, \porb FAP vs the stellar axial inclination used to recover the starspot contribution, $i_{\rm rec}$. The input simulation of $i = 60\degs$ is indicated with the vertical dashed line.}
\protect\label{fig:faps}
\end{center}
\end{figure}

\section{Simulations}
\protect\label{section:sims}

We investigate whether for typical M stars with significant rotation (i.e. $\vsini \leq$ 10 \kms) and un-evolving starspot patterns, we are able to distinguish between both starspot signals and stellar reflex motion due to an orbiting planet with modest observational effort. The issue of starspot evolution is discussed further in \S \ref{section:discussion}. We first begin by simulating a specific case in order to demonstrate the feasibility of the technique.

\subsection{Proxima Centauri analogue model}
\protect\label{section:proxcenmodel}

We illustrate the procedure by adopting a model with the stellar parameters of Proxima Centauri and the orbital period of Proxima \hbox{Centauri b} reported in \citet{anglada16proxima}. While Proxima Centauri possesses a very low \vsini\ \hbox{$< 0.1$\ \kms}, we investigate projected equatorial rotation velocities in the \hbox{\vsini\ $= 1$\,-\,$10$\ \kms} range following the finding that fully convective M dwarfs are on average moderate rotators \citep{jenkins09mdwarfs}. We have chosen to simulate a planetary RV signal with a \hbox{\porb $= 11.2$ d} orbit that induces a stellar semi-amplitude of \hbox{\kstar $ = 2$\ \ms} (slightly higher than the 1.4 \ms\ reported for Proxima Centauri b). The model parameters are summarised in \hbox{Table \ref{tab:params}}. In this section we use an axial inclination of $i = 60$\degs\ and projected equatorial rotation velocity of \vsini = \hbox{5 \kms}. A rotation period of \prot $= 1.23$ d was calculated from the estimated \rstar $= 0.14$\rsun of Proxima Centauri using \hbox{\prot $=\ 2 \pi R_*\ {\rm sin}\ i / v\ {\rm sin}\ i$}.

\subsection{Simulated line profiles and planet recovery}
\protect\label{section:simulatedlines}

\subsubsection{S/N and number of observations}
\protect\label{section:snepochs}

The maximum entropy procedure requires high S/N ratios to enable recovery of spot features from the line distortions. This can be achieved either by considering many lines simultaneously or by applying least squares deconvolution \citep{donati97zdi} to typically several thousand lines in each spectrum to obtain a single high S/N ratio line profile. Deconvolution has a computational benefit compared with simultaneous fitting of multiple lines. Unlike a straightforward cross-correlation profile, the least squares deconvolved profile accounts for line blending effects. We applied our implementation of least squares deconvolution \citep{barnes98aper,barnes12rops} to the 0.64\,-\,1.03 \micron\ region of M dwarf spectra { (observed with {\sc uves})} with mean S/N ratios of $25 \lesssim$ \snrobs $\lesssim 140$. This yielded deconvolved profiles with $2000 \lesssim$ \snrdecon $\lesssim 12,000$ { and a mean \snrdecon $\sim 5300$} (see \hbox{table 1} of \citealt{barnes14rops}). { Although \snrobs in the above range can be achieved for the brightest mid-M dwarfs in the reddest orders with \harps, red-optical and near infrared surveys are required to enable sufficient wavelength coverage to achieve \snrdecon $\geq 2000$\ on a large sample of mid-late M dwarf targets.}

The simulations presented here consider typical line profile \snrdecon ratios of $2000$, $5000$ and $10000$ that would be expected from spectra with respective mean \snrobs $\sim$ $25$, $60$ and $120$. We used the least squares deconvolved profile of the slow rotator, GJ 1061, observed with spectral resolution, $R$\,$\sim$\,100,000 \citep{barnes14rops}, to represent the unbroadened local intensity profile during modelling. Line profiles with starspot asymmetries and an additional sinusoidal velocity variation specified by the planet observables, \porb and \kstar (i.e. a circular orbit is assumed), were generated with \dotstom assuming one observation per night for \hbox{\nobs = $15$}, $30$ and $60$ nights. A \hbox{3 hour} random time shift on each observation was applied to minimise day aliases. 

\subsubsection{Radial Velocities}
\protect\label{section:rvs}

The simulated line profiles containing a planet RV signal and line profile distortions due to spots were cross-correlated using the mean profile (i.e. sum of all profiles) as a reference. For the Proxima Centauri analogue model, with \snrdecon = 5000, the resulting simulated radial velocities, \rvsim, are shown by the red triangles in Fig. \ref{fig:simulations1} (left panels) using Models 2, 3 and 4 with 15 (top), 30 (middle) and 60 nights (bottom) of observations. For each data set, we used \dotstom to fit the simulated line profiles, but without allowing for the sinusoidal variation of the line centroids due to an orbiting planet (i.e. in contrast to \citealt{petit15planets} who simultaneously recovered the planet RVs in the presence of spots). The simulated line profiles were fit using the input parameters, $\vsini$, $i$ and \prot. The fitted or recovered line profiles thus only contain the line profile distortions due to starspots. These were then cross-correlated against the mean recovered line profile to obtain the recovered velocities, \rvrec, shown by the green dashed lines in Fig. \ref{fig:simulations1} (left panels). 
For Models 3 and 4 the line asymmetries induced by the spots are the dominant contributor to the RVs. Since the spots in Model 2 are too small to be resolved, the fitting procedure is not able to model the line profiles adequately. As a result, for Model 2, the \rvrec points do not well match the \rvsim points. The corrected radial velocities are obtained by subtracting the recovered RVs from the simulated RVs, where \hbox{\rvcorr = \rvsim - \rvrec} (blue circles). The middle panels of Fig. \ref{fig:simulations1} show the recovered RVs (\rvrec) vs the simulated RVs (\rvsim) showing a linear trend in all instances. The shallow slope for Model 2 is a reflection of the inability of the modelling process to recover the small spots. 

\subsubsection{Periodograms and false alarm probabilities}
\protect\label{section:periodograms_faps}

The right panels of Fig. \ref{fig:simulations1} show the periodograms of the simulated RVs and the corrected RVs, \rvsim and \rvcorr, for each simulation. Tick marks indicate the simulated periods, \prot $= 1.23$\ d and \porb $= 11.2$\ d and 0.1, 1 and 10 per cent false alarm probability (FAP) levels are shown. A number of additional peaks arising from the window function sampling and aliasing are present (e.g. \prot/2 = 0.615 d peak and the 1 d sampling beat period with \prot at 5.3 d). However, it is noteworthy that the solar maximum activity levels from Model 2 are not sufficient to mask the \porb $= 11.2$\ d period, even with only 15 observations. This is expected since in Fig. 3 of \citet{barnes11jitter}, we showed that for the low-contrast case, $< 1$ \ms\ jitter is expected from Model 2 for \vsini\ = 5 \kms in the absence of limiting photon noise. For Models 3 and 4, the \porb peak only appears at low significance in the periodograms of the simulated RVs (red dashed curves). For the Model 3 and 4 spot corrected \rvcorr data, the \prot peak is no longer present in the periodograms (blue { solid} curves). At the same time, the \porb peak significance is boosted. Other peaks that coincide with beating between the 1 day sampling and \porb however appear at close to $\sim 1$\ d, with $\sim 2$\ d aliases. In the case of the \nobs = 60 epoch simulation, these peaks are of comparable or greater significance than \porb. Thus, while we can effectively remove the spot jitter using only relatively few observations, the \porb $= 11.2$ d peak is still of low significance, reaching only $\gtrsim$ 10 per cent FAP for Models 2 and 3 with 30 observations. Fig. \ref{fig:faps} (left panel) shows that for our simulated models with \hbox{\vsini =} \hbox{$5$ \kms}, \porb is recovered with $< 0.1$ per cent FAP for Models 2 and 3 with \nobs $\sim 54$, while \hbox{\nobs = 58} is sufficient to recover \porb with $< 1$ per cent FAP for Model 4.

\subsection{Identification of the stellar rotation period}
\protect\label{section:rotationperiod}

\begin{figure}
\begin{center}
\vspace{-2mm}
\includegraphics[width=49mm,angle=270]{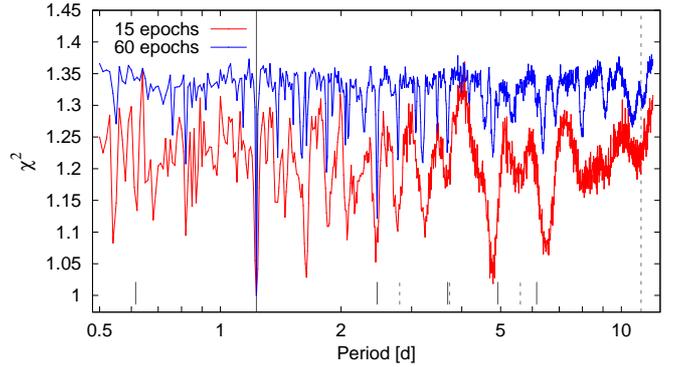}  \\
\caption{Stellar rotation period recovery with DoTS for Model 3 with 15 (red/solid) and 60 (blue/dotted) observations. \chisq normalised to the minimum recovered value is plotted vs stellar rotation period. \prot = 1.23 d is shown by the solid vertical line. Solid tick marks denote \prot$/2$, $2$\prot, $3$\prot and $4$\prot. The dashed vertical line and tick marks indicate \porb \hbox{$= 11.2$ d}, \hbox{\porb/$2$}, \hbox{\porb/$3$} and \hbox{\porb/$4$}.}
\protect\label{fig:starsperiod}
\end{center}
\end{figure}

The recovered spot RVs in \S \ref{section:rvs}, were obtained by fitting the simulated line profiles with the simulated (i.e. known) rotation period, \prot. However, identification of the rotation period is not necessarily straightforward; a \prot $= 1.23$ d peak is not always significant, or the most significant peak in the periodograms in Fig. \ref{fig:simulations1} (red/dashed curves). The \hbox{\prot $= 1.23$ d} peak strength depends on \nobs and the number of spots. The window function, aliasing and possible beat periods result in peaks at other periods (see \S \ref{section:simulatedlines} above). Rather than relying on Lomb-Scargle periodogram analysis, we used \dotstom to search for the simulated stellar rotation period. Since maximum entropy regularised line fitting is an iterative process, the \chisq achieved after a fixed number of iterations can be used as a criterion for recovering the best fitting parameters. This procedure works well for the stellar rotation period since spots are periodically visible at the same location on the star as viewed by an observer. Fig. \ref{fig:starsperiod} shows that even with a 15 epoch simulation, \hbox{\prot =} \hbox{1.23} d is recovered with the minimum \chisq (i.e. proportional to the maximum likelihood). With \nobs = 60, confidence in the period is increased relative to other local \chisq minima. \prot can thus be determined with a modest number of observations. Because the periodicities are determined directly from comparing information in the line profiles with a physical model, it is possible to recover periodicities more confidently than with a straightforward Lomb-Scargle periodogram search for sinusoidal signals. Nevertheless, using an armoury of diagnostic tools, including periodogram analyses (i.e. Fig. \ref{fig:simulations1} generally shows a peak at \prot) and photometry, will offer the most effective means of recovering \prot.

\subsection{Rotation and Stellar inclination}
\protect\label{section:inclination}

In addition to recovering \prot, the projected equatorial rotation velocity, \vsini, must be recovered. It is common practice in Doppler imaging to optimise \vsini, which correlates with line strength when using a fixed line profile to represent the local stellar spectrum (e.g. see \citealt{barnes00pztel} and \citealt{cameron01mapping}). This procedure is often needed for each data set to optimise the fit and minimise image artefacts. These parameters are generally easily determined by \chisq minimisation  and are orthogonal to periodic signatures. In addition, the stellar axial inclination, $i$, must be specified during the starspot modelling and line profile recovery. The stellar radius cannot generally be measured directly from observation, which means that $i$ is not generally known. We performed a \chisq search for a range of stellar axial inclinations, \irec, to determine the effect of an incorrectly determined axial inclination. A minimum \chisq is always attained at the simulated combination of \vsini and \prot, irrespective of adopted axial inclination for $30$\degs\ $\leq$ \irec $\leq 90$\degs. These parameters are therefore effectively independent of choice of \irec. However for the trial \irec values, only 2 per cent r.m.s. variation in $\chi^2_{\rm min}$ was seen, with no discernible trend in \chisq vs \irec in the $30$\degs-\,$90$\degs\ range above the scatter. Thus, although the simulated inclination of $i = 60$\degs\ cannot be recovered for \vsini = 5 \kms, this does not preclude identification of \vsini and \prot, which are correctly recovered at all trial \irec values.

A robustly determined rotation period, \prot and \vsini still enable $i$ to be estimated for an assumed stellar radius. The typical 10 per cent uncertainty for M dwarf estimates of \rstar \citep{stassun12radii,williams15}, will dominate uncertainty in the axial inclination determination. For instance, uncertainties of 10 per cent in ${\rm sin}\ i$ yield $i = 30^{+3.4}_{-3.3}$\degs, $45^{+6.1}_{-5.5}$\degs, $60^{+12.3}_{-8.8}$\degs, $75^{+15}_{-9.6}$\degs\ and $90^{+0}_{-25.8}$\degs\ (upper limits at 75\degs\ and 90\degs\ arise from requiring sini\ $i \leq 1$).
For \rvsim data using $i = 60$\degs\ and \nobs = 60, Fig. \ref{fig:faps} (right) shows the FAP for \rvcorr data points using $i = 30$\degs-\,$90$\degs\ (in $15$\degs\ steps). In fact, there is a tendency towards improved FAPs when using higher inclinations (i.e. \irec = $75$\degs\ and $90$\degs) for recovery. This is likely related to the lack of resolution elements, when instrumental resolution and stellar \vsini are of comparable magnitude. 

\section{Sensitivities}
\protect\label{section:sensitivities}

\begin{figure}
\begin{center}
\includegraphics[width=100mm,angle=270]{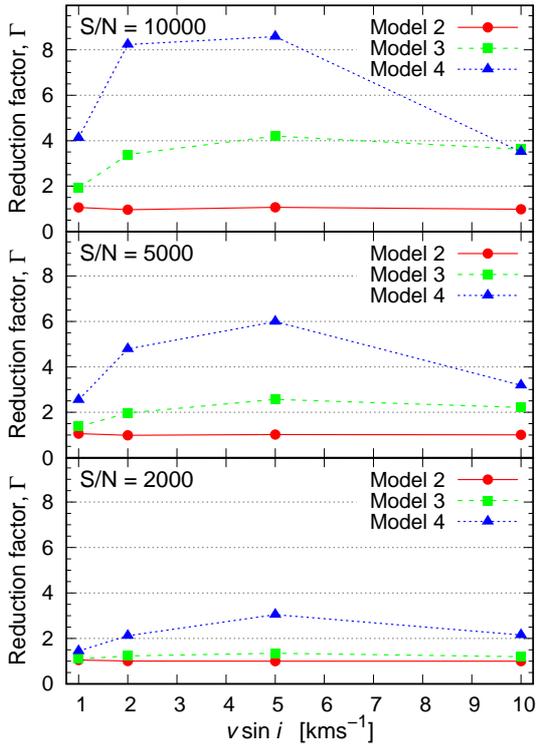}  \\
\caption{Starspot noise reduction factor, $\Gamma$, as a function of \vsini for S/N = $2000$, $5000$ and $10000$ (bottom to top).}
\protect\label{fig:mitigation}
\end{center}
\end{figure}

As every object and data set combination is unique, a full exploration of parameter space is beyond the scope of this paper. Nevertheless, a more complete picture of the range of sensitivities can be gained by exploring the S/N and \vsini values that are expected empirically. Fig. \ref{fig:mitigation} shows the reduction in starspot noise achieved, $\Gamma$ = $\rho_{\rm sim} / \rho_{\rm corr}$ (where $\rho_{\rm sim}$ and $\rho_{\rm corr}$ are the r.m.s. for the \rvsim and \rvcorr data points), as a function of \vsini for the simulated S/N ratios. The lowest values of $\Gamma$ are found for Model 2, where $\Gamma \sim 1$. Again this is due to an inability to recover the small spots, and in some combinations, the noise may be marginally increased by $\sim 5$ per cent. The procedure is most efficient for \hbox{\vsini $=$} \hbox{$5$ \kms}, where a maximum \hbox{$\Gamma =$\ $8.6$} is achieved. At \hbox{\vsini $=$} \hbox{$1$\ \kms}, the efficiency is generally lower, as fitting distortions below the instrumental resolution is less effective. For \hbox{\vsini $=$} \hbox{$10$ \kms}, although starspot distortions are modelled well, the broader lines due to higher \vsini limit the efficiency of the procedure by increasing the RV uncertainties.

\begin{figure}
\begin{center}
\vspace{-3mm}
\includegraphics[width=105mm,angle=270]{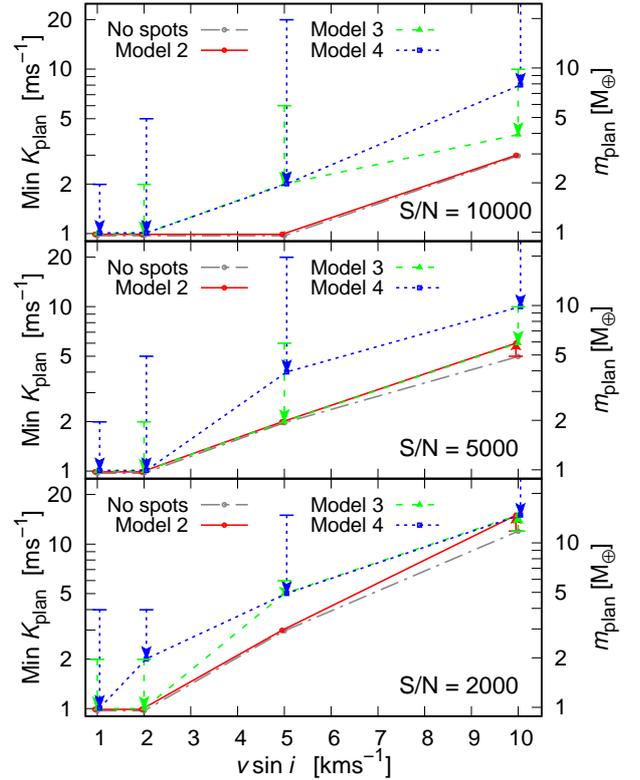}  \\
\caption{Minimum detected amplitude and planet mass with FAP $\leq 0.001$ ($0.1$\%) for S/N = 2000, 5000 and 10000 (bottom to top). For Models 2, 3 and 4, the minimum detected \kstar and equivalent planet mass, $m_{\rm plan}$, is plotted for the simulated \vsini values after applying spot correction (with an assumed $i = 60$\degs, \kstar/$m_{\rm plan}$ $\sim 1$ for the adopted \mstar and \porb). The vertical arrows indicate the change in sensitivity achieved by subtracting the fitted starspot radial velocities. Before correction, the sensitivity limits for \hbox{\vsini =} \hbox{$10$ \kms}\ are \hbox{\kstar $= 50$} \ms (i.e. \mplan $\sim$\,50 \mearth). The grey dot-dashed line is the precision limit for an un-spotted star.}
\protect\label{fig:fapsimulation}
\end{center}
\end{figure}

For projected equatorial rotation velocities, \hbox{\vsini = $1$}, $2$, $5$\ and\ \hbox{$10$ \kms}, we expect \prot $= 6.13$, $3.10$, $1.23$ and \hbox{$0.613$ d} for a $0.14$\rsun star with $i$ = $60$\degs. With \porb fixed at \hbox{$11.2$ d}, we have estimated the minimum \kstar recovered with $\leq 0.1$ FAP using \nobs = 60 before and after subtracting the fitted starspot contributions. A noise floor of 1 \ms is assumed and we simulated recovery FAPs for a number of discrete values of \kstar up to 50 \ms. The increase in sensitivity with increasing S/N ratio and decreasing \vsini that one intuitively expects is seen in Fig. \ref{fig:fapsimulation}. The connected points indicate the sensitivity for each model after removing the spot contribution. The vertical arrows indicate the change in sensitivity from the \rvsim points to the \rvcorr points. The limiting precision is also shown by the grey dot-dashed line, and is derived from line profiles generated with an unspotted model. The absorption lines are more effectively cleaned for Model 4 since these spots induce the largest amplitudes and are most easily fit. The procedure is ineffective for Model 2 because the spots are not resolved. For Model 2, when the S/N ratio is low and \vsini is 10 \kms, the modelling adds noise and decreases the sensitivity slightly.

{ For the fixed \porb = \hbox{$11.2$ d}}, in the case of the slowest rotators with \hbox{\vsini $=$} \hbox{$1\,-\,2$ \kms}, \nobs = $60$ should enable detection of \hbox{$1\,-\,2$ \mearth} planets, and is limited to \hbox{$2$ \mearth} only for \hbox{S/N =} $2000$. For \hbox{\vsini =} \hbox{$5$ \kms} and the highest \snrdecon = $10000$, \hbox{$1$ \mearth} planets can potentially still be detected if \hbox{Model $2$} spot levels are realistic. More typical limits of \hbox{$2$\,-\,$4$ \mearth} are achieved for \hbox{Models $3$} and $4$ after removal of spot RVs. Once \vsini reaches \hbox{10 \kms}, only \hbox{$\geq 4$ \mearth} planets are likely to be detected for the simulated observations, assuming Models $3$ and $4$ are representative of spot coverage. 

\section{Summary \& Discussion}
\protect\label{section:discussion}

Our ability to recover and remove stellar activity induced radial velocities is demonstrated for late-M dwarfs with distributed starspot patterns. We have not attempted a complete exploration of parameter space. Nonetheless we are confident that a typical habitable zone planet, bounded by the runaway and maximum greenhouse conditions, with $a =$ $0.042$\,-\,\hbox{$0.082$ AU} and \porb $\sim$\,$9.1$\,-\,$24.5$ d \citep{delfosse00masslum,kopparapu13hz,kopparapu13hzcorr}, could be detected orbiting an M6V analogue of Proxima Centauri \citep{anglada16proxima}. The size of the line distortions induced by the spots, the resolving power and the S/N of the observed spectra are the most important factors that determine the efficacy of the method. Our adopted spot models are extrapolated solar spot size distributions, with spot contrasts and surface distributions based on observations of rapidly rotating M dwarfs, although we do not know the exact form of the spot distributions for moderately rotating, fully convective M dwarfs. Nevertheless, with reasonable levels of spot coverage, we are confident spot activity can be reduced to a level that enables 1 \mearth signals to be detected in spectra observed with typical \snrobs $\sim$\,$60$ (corresponding to the simulated deconvolved profiles with \snrdecon $\sim$\,5000 in this paper). 

Observations that aim to detect planets and determine planet occurrence rates for fully convective M dwarfs will only be unbiased if moderate rotators are surveyed. This means removing spots is crucial. We find that for moderately spotted stars, 2 \mearth planets can be detected for \vsini = 5 \kms, while 4 \mearth planets can be detected at \vsini = 10 \kms. Without corrections, the mass limits are up to a factor of $\Gamma = 8.6$ higher in the I band. Applying the technique to M dwarfs with \hbox{\vsini $> 5$ \kms} at \snrobs $\leq 25$ (yielding \snrdecon = 2000) will not necessarily be advantageous if spot coverage is modest (i.e. Models 2 and 3 with filling factors of $< 2$ per cent). Clustering of spots in groups potentially results in larger amplitude starspot distortions (i.e. where spots within a spot group are not resolved in the line profile), which are more easily fit. Although Doppler images of rapidly rotating M dwarfs (i.e. with \vsini $\geq 20$ \kms) reveal distributed starspot patterns, it is less clear whether spot activity in stars with \vsini $\leq 10$ \kms will show similar distributed spot patterns, or spots located in active regions, as seen on the Sun and other earlier G and K dwarfs.

Obtaining rotation periods for the latest M stars is troublesome owing to the low contrast and distributed nature of the spots. As demonstrated in \S \ref{section:rotationperiod}, identification of the planetary period from other short-period alias peaks in Fig. \ref{fig:simulations1} is likely to require careful consideration of a combination of photometry, activity parameters and line diagnosis tools. We have shown that using maximum entropy fitting offers one such method for reliable determination of the rotation period,  where periodogram analysis alone does not yield an unambiguous peak. Further, for real observations with additional systematics, traditional Lomb-Scargle periodogram analyses do not offer the optimal path to recovery of planet-induced periodicities. Frequentist and Bayesian approaches that incorporate the stellar signal into the period analysis \citep{baluev13gj581,tuomi14mdwarfs,anglada16proxima}, using additional instrumental and atmospheric noise priors, will provide a better means of assessing the periodicities. 

For slowly rotating stars, where \vsini $<<$ $R$ (the instrumental resolution), the maximum entropy method presented here will not be effective. For \vsini $< 1$ \kms even if spot distributions similar to Model 3 or 4 were seen, it is unlikely that they will be resolved since micro- and macro-turbulence will dominate the local intensity profile width. Additional tools for assessing line shape behaviour can also be applied to assess correlations with instrumental systematics and activity \citep{berdinas16harps,anglada16proxima}. For instance, although late-M field stars typically exhibit significant rotation, Proxima Centauri, with a probable \hbox{\vsini $<$ 0.09 \kms}, has demonstrated line profile \citep{anglada16proxima} variability on an $\sim$$80$ d time scale, matching the photometric modulation. The origin of the variability in M dwarf stars is also not clear. The variability in the second central moment (essentially the width of the spectral lines), was shown by \citet{anglada16proxima} to be anti-correlated with the photometry. In other words, the line width is greatest when the stellar flux is lowest. This implies potential effects from plage regions, which are likely associated with spot activity, and which will also modify the line equivalent width. Further modelling to enable recovery of line variations due to plage may therefore be appropriate as an extension to the assumed two-temperature (photosphere + spot) simulations that we present in this work. 

Recovery of surface temperature inhomogeneities will be further complicated by the fact that starspots are not static but constantly evolving with time. Spot groups on dwarf stars are nevertheless stable on timescales of months \citep{barnes98aper,bradshaw14spotlife}, while coherence of individual spots on day-week timescales has been used to measure differential rotation (e.g. \citealt{barnes05diffrot,cameron02twist}). This is also true for fully convective active M dwarfs, which show spot stability over several days \citep{barnes16zenodo}. { Differential rotation with small amplitude or consistent with solid body rotation has been found by \citet{morin08v374peg,morin08mdwarfs} and \citet{barnes16zenodo}} in these targets, { and} may be responsible for stability of spots on even longer timescales.
Radial velocity surveys that target stars intensively, with observations obtained in relatively short campaigns are thus likely to be the most successful. The campaign that identified Proxima Centauri b \citep{anglada16proxima} obtained $\sim$\,$60$ observations during a two month timespan. On short timescales, peak aliasing is minimised, while stellar activity variability can more easily monitored and accounted for. This contrasts with typical multi-epoch surveys that often obtain a few observations per season over several years. We have demonstrated that the Doppler imaging method works well, even with only 15 observations that sample the range of rotation phases, at relatively high cadence over a short observing baseline. Splitting a 60 epoch campaign into subgroups of 15 observations would enable the effects of spots to be mitigated, while further minimising the effects of spot evolution.

The simulations presented indicate that for stars with distributed starspots at low contrast, removal of starspot induced radial velocities is challenging. Nevertheless, the technique presented in this paper demonstrates that we can model and correct for starspot induced radial velocities in stars with \vsini = 1\,-10 \kms. With higher spot contrasts, the technique can be applied to less active and more slowly rotating stars. Since no prior assumptions about the spot distributions are required, radial velocity surveys aimed at detecting earth-mass planets at all spectral types will benefit from this technique. We have clearly demonstrated that high cadence observations on the timescale of the stellar rotation period are essential for reliable RV detection of planets orbiting active stars. 

\section*{Acknowledgments}
{ We thank the anonymous referee for taking the time to review this manuscript.} J.R.B. and C.A.H. were supported by the STFC under the grant ST/L000776/1. S.V.J. acknowledges research funding by the Deutsche Forschungsgemeinschaft (DFG) under grant SFB 963/1, project A16. JSJ acknowledges funding by Fondecyt through grants 1161218 and 3110004, and partial support from CATA-Basal (PB06, Conicyt), the GEMINI-CONICYT FUND and from the Comit\'{e} Mixto ESO-GOBIERNO DE CHILE. 






\bsp	
\label{lastpage}
\end{document}